\documentclass[aps,prc,preprint,showpacs,showkeys]{revtex4}
\makeatletter

\newcommand{\Rmnum}[1]{\expandafter\@slowromancap\romannumeral #1@}
\makeatother
\usepackage{amssymb}
\usepackage{amsmath}
\usepackage{graphicx}
\usepackage{bm}
\usepackage{epsf}

\begin{document}

\title{Influence of the symmetry energy on nuclear
 \textquotedblleft pasta\textquotedblright\ in neutron star crusts}

\author{S. S. Bao}
\affiliation{School of Physics, Nankai University, Tianjin 300071, China}
\author{H. Shen}
\affiliation{School of Physics, Nankai University, Tianjin 300071, China}
\email{shennankai@gmail.com}

\begin{abstract}
We investigate the effects of the symmetry energy on nuclear
\textquotedblleft pasta\textquotedblright\ phases
and the crust-core transition in neutron stars. We employ the
relativistic mean-field approach and the coexisting phases method
to study the properties of pasta phases presented in the inner crust
of neutron stars. It is found that the slope of the nuclear symmetry
energy at saturation density plays an important role in the
crust-core transition and pasta phase properties.
The correlation between the symmetry energy slope and the crust-core
transition density obtained in this study is consistent
with those obtained by other methods.
\end{abstract}



\maketitle

\section{Introduction}
\label{sec:1}

Neutron stars are fascinating laboratories for the study of cold matter
ranging from subnuclear density to several times normal nuclear matter
density. It is generally believed that a neutron star has an outer crust of
nuclei in a gas of electrons, an inner crust of nuclei in a gas of neutrons
and electrons, and a liquid core of homogeneous nucleonic
matter~\cite{Webe05,PR00,PR07}.
The inner crust is of great interest because of its importance in
astrophysical observations and its complex phase structure~\cite{Stei98}.
Nuclei with exotic shapes, known as pasta phases, are expected to be
present in the inner crust of neutron stars~\cite{Rave83,Wata00,Oyam07}.
The stable nuclear shape may change from droplet to rod,
slab, tube, and bubble. At the crust-core transition density,
the pasta nuclei are dissolved into uniform matter since
the homogeneous phase has a lower free energy than the pasta phase.
In past decades, the existence of pasta phases has been studied using
various methods, such as the liquid-drop model~\cite{Wata00} and the Thomas-Fermi
approximation~\cite{Oyam07,Mene08,Mene10,Gril12}.
The authors of Refs.~\cite{Mene10,Gril12} investigated the properties
of pasta phases within the Thomas-Fermi approximation, and they found
that the internal structure of pasta phases and the interface with
the homogeneous phase are sensitive to the models used.
The differences in the symmetry energy among various models
have a significant impact on the pasta phases and the crust-core
transition~\cite{Gril12,Duco11}.

The density dependence of the nuclear symmetry energy is very important for
understanding many phenomena in both nuclear physics and
astrophysics~\cite{PR07,LiBA08}.
The symmetry energy $E_{\rm sym}$ at saturation density
is constrained by experiments to be around $30\pm 4$ MeV,
while the slope of the symmetry energy $L$ at saturation density
is still rather uncertain and may vary from about $20$
to $115$ MeV~\cite{Chen13}.
The effect of the symmetry energy on neutron star properties has
been studied by many authors~\cite{Oyam07,Duco11,Mene11}.
In Ref.~\cite{Duco11}, the influence of the symmetry energy on the crust-core
transition was studied using a variety of nuclear effective
models, in which the crust-core transition density obtained from
the dynamical and thermodynamical methods decreases clearly with increasing $L$.
This correlation between the transition density and $L$
has also been discussed in many works~\cite{Oyam07,LiBA08,Gril12}.
In Ref.~\cite{Oyam07}, the properties of nuclei in the inner crust were fund
to be sensitive to the density dependence of the symmetry energy
using a parametrized Thomas-Fermi approach.
The pasta phases were also studied within a self-consistent
Thomas-Fermi approach in Ref.~\cite{Gril12}, and it was found
that $L$ could have dramatic effects on the pasta structure.
It is interesting and important to explore clear correlations
between $L$ and the crust-core transition.

In this article, we systematically examine how the crust-core transition
and pasta phase properties depend on the symmetry energy slope $L$ using the
coexisting phases method. We employ the Wigner-Seitz approximation to
describe the inner crust matter, in which the conditions of $\beta$
equilibrium and charge neutrality are satisfied. The properties of
the Wigner-Seitz cell can be obtained from the energy minimization
for zero temperature matter.
In the coexisting phases method, the matter inside the Wigner-Seitz
cell separates into a dense phase and a dilute phase which satisfy
Gibbs conditions for phase equilibrium.
We use the relativistic mean-field (RMF) approach to describe
these two phases, while the surface tension is properly determined
by a Thomas-Fermi calculation.
The RMF theory has been successfully used to study various phenomena
in nuclear physics over the past decades~\cite{Sero86,Ring90,Meng06}.
The RMF theory has recently been reinterpreted by the relativistic Kohn-Sham
density functional theory, which has been widely employed
in the treatment of the quantum many-body problem in atomic, molecular,
and condensed matter physics. In the RMF approach, baryons interact
through the exchange of isoscalar scalar and vector mesons
($\sigma$ and $\omega$)
and an isovector vector meson ($\rho$), while the parameters
are generally determined by fitting to some nuclear matter properties
or ground-state properties of finite nuclei.
In the present work, we employ two different RMF models, namely the TM1~\cite{TM1}
and IUFSU~\cite{IUFSU} parametrizations, so that we can examine the model
dependence of the results obtained. It is known that both TM1 and IUFSU models
can well reproduce the ground state properties of finite nuclei,
including unstable nuclei, and a maximum neutron-star mass $\sim 2 M_\odot$.
The TM1 model has been successfully used to construct the equation of state
for supernova simulations and neutron stars~\cite{Shen02,Shen11},
while the IUFSU model was proposed to overcome a smaller neutron-star mass
predicted by the FSU model~\cite{IUFSU}.
These two models include nonlinear terms for both $\sigma$ and $\omega$
mesons, while the IUFSU model includes an additional
$\omega$-$\rho$ coupling term. It has been extensively discussed
that the $\omega$-$\rho$ coupling plays an important
role in modifying the density dependence of the symmetry energy and affecting
the neutron star properties~\cite{Mene11,Gril12,Horo01,Horo03,Prov13}.
In order to examine the influence of $L$ on pasta phase
properties, we generate two sets of models based on the TM1 and IUFSU parametrizations.
In one set of models, all saturation properties are the same
except the symmetry energy slope $L$, which is controlled by tuning
the $\omega$-$\rho$ coupling strength.
By using the set of models with different $L$, it is possible to study
the impact of $L$ on the crust-core transition and
pasta phase properties.

This article is arranged as follows. In Sec.~\ref{sec:2},
we present the formalism used in the present work.
In Sec.~\ref{sec:3}, we show the numerical results and discuss
the influence of $L$ on pasta phase properties and
possible correlations between $L$ and the crust-core transition.
Section~\ref{sec:4} is devoted to the conclusions.

\section{ Formalism}
\label{sec:2}

We adopt the Wigner-Seitz approximation and the coexisting phases method
to describe the inner crust matter, in which the conditions of $\beta$
equilibrium and charge neutrality are satisfied.
The matter inside the Wigner-Seitz cell separates into a dense phase
and a dilute phase. We use the RMF theory to describe these two phases.
In the RMF approach, baryons interact through the exchange of isoscalar
scalar and vector mesons ($\sigma$ and $\omega$) and an isovector vector
meson ($\rho$). We employ the TM1~\cite{TM1} and IUFSU~\cite{IUFSU}
parametrizations of the RMF models,
which are known to be successful in reproducing the ground state properties
of finite nuclei including unstable nuclei.
The nucleonic part of the Lagrangian density takes the form
\begin{eqnarray}
\label{eq:LRMF}
{\cal L}_{\rm{RMF}} & = & \bar{\psi}\left[i\gamma_{\mu}\partial^{\mu}
-\left(M+g_{\sigma}\sigma\right)-\left(g_{\omega}\omega^{\mu}
+\frac{g_{\rho}}{2}\tau_a\rho^{a\mu}\right)\gamma_{\mu}
\right]\psi  \nonumber\\
 && +\frac{1}{2}\partial_{\mu}\sigma\partial^{\mu}\sigma
-\frac{1}{2}m^2_{\sigma}\sigma^2-\frac{1}{3}g_{2}\sigma^{3}
-\frac{1}{4}g_{3}\sigma^{4} \nonumber\\
 && -\frac{1}{4}W_{\mu\nu}W^{\mu\nu}
+\frac{1}{2}m^2_{\omega}\omega_{\mu}\omega^{\mu}
+\frac{1}{4}c_{3}\left(\omega_{\mu}\omega^{\mu}\right)^2   \nonumber\\
 && -\frac{1}{4}R^a_{\mu\nu}R^{a\mu\nu}
+\frac{1}{2}m^2_{\rho}\rho^a_{\mu}\rho^{a\mu}
+\Lambda_{\textrm{v}} \left(g_{\omega}^2 \omega_{\mu}\omega^{\mu}\right)
\left(g_{\rho}^2\rho^a_{\mu}\rho^{a\mu}\right),
\end{eqnarray}
where $\psi$ is an isodoublet nucleon field.
$\sigma$, $\omega^{\mu}$, and $\rho^{a\mu}$ are
$\sigma$, $\omega$, and $\rho$ meson fields with masses
$m_{\sigma}$, $m_{\omega}$, and $m_{\rho}$.
$W^{\mu\nu}$ and $R^{a\mu\nu}$ are the antisymmetric field tensors
for $\omega^{\mu}$ and  $\rho^{a\mu}$, respectively.
It is known that the inclusion of nonlinear $\sigma$ terms is
essential to reproduce the properties of nuclei quantitatively and provides
a reasonable value for the incompressibility, while the nonlinear $\omega$
term is added to reproduce the density dependence of the nucleon self-energy
obtained in the relativistic Brueckner-Hartree-Fock theory~\cite{TM1}.
We include the $\omega$-$\rho$ coupling term as described in~\cite{IUFSU},
which is essential in modifying the density dependence of the symmetry energy.

In the RMF approximation, the meson fields are treated as classical fields
and the field operators are replaced by their expectation values.
For a static system, the nonvanishing expectation values
are   $\sigma =\left\langle \sigma    \right\rangle$,
      $\omega =\left\langle \omega^{0}\right\rangle$,
and   $\rho   =\left\langle \rho^{30} \right\rangle$.
The energy density of homogeneous nuclear matter
can be written as
\begin{eqnarray}
\varepsilon &=&\sum_{i=p,n}\frac{1}{\pi^2}
     \int_{0}^{k^{i}_{F}}{\sqrt{k^2+{M^{\ast}}^2}}k^2dk   \nonumber \\
&& + \frac{1}{2}m^2_{\sigma}{\sigma}^2+\frac{1}{3}{g_2}{\sigma}^3
     +\frac{1}{4}{g_3}{\sigma}^4  \nonumber  \\
&& + \frac{1}{2}m^2_{\omega}{\omega}^2+
     \frac{3}{4}{c_3}{\omega}^4
     + \frac{1}{2}m^2_{\rho}{\rho}^2
     + 3{\Lambda}_{\textrm{v}}\left(g^2_{\omega}{\omega}^2\right)
     \left(g^2_{\rho}{\rho}^2\right),
\end{eqnarray}
and the pressure is given by
\begin{eqnarray}
P &=& \sum_{i=p,n}\frac{1}{3\pi^2}\int_{0}^{k^{i}_{F}}
      \frac{1}{\sqrt{k^2+{M^{\ast}}^2}}k^4dk    \nonumber  \\
&& - \frac{1}{2}m^2_{\sigma}{\sigma}^2-\frac{1}{3}{g_2}{\sigma}^3
     -\frac{1}{4}{g_3}{\sigma}^4     \nonumber \\
&& + \frac{1}{2}m^2_{\omega}{\omega}^2
      +\frac{1}{4}{c_3}{\omega}^4
      + \frac{1}{2}m^2_{\rho}{\rho}^2 +
      \Lambda_{\textrm{v}}\left(g^2_{\omega}{\omega}^2\right)
      \left(g^2_{\rho}{\rho}^2\right),
\end{eqnarray}
where $M^{\ast}=M+g_{\sigma}{\sigma}$ is the effective nucleon mass.
With the TM1 and IUFSU parameter sets listed in Table~\ref{tab:1},
we can achieve similar values for the saturation density and binding energy per nucleon,
but different results for the symmetry energy~\cite{TM1,IUFSU}.
The symmetry energy $E_{\textrm{sym}}$ is given by
\begin{equation}
E_{\textrm{sym}}=\frac{1}{2}{\left[\frac{\partial^2\left(\varepsilon/n_b\right)}
{\partial\alpha^2}\right]}_{\alpha=0}=
\frac{k^2_F}{6\sqrt{k^2_F+{M^\ast}^2}}+\frac{g^2_\rho}{8{m^{\ast}_{\rho}}^2}n_b,
\end{equation}
where $n_b$ is the baryon number density,
$\alpha=\left(n_n-n_p\right)/n_b$ is the asymmetry parameter,
and ${m^{\ast}_{\rho}}^2=m^2_{\rho}+2\Lambda_{\textrm{v}}g^2_\rho
{g^2_\omega}{\omega}^2$.
The slope of the symmetry energy $L$ at saturation density $n_0$ is given by
\begin{equation}
L=3n_0\left[\frac{\partial E_{\textrm{sym}}\left( n_b \right)}
{\partial{n_b}}\right]_{n_b=n_0}.
\end{equation}
We note that $E_{\textrm{sym}}=36.9$ MeV and $L=110.8$ MeV are obtained for TM1, while
$E_{\textrm{sym}}=31.3$ MeV and $L=47.2$ MeV are obtained for IUFSU.

In order to study the effect of $L$ clearly, we generate two sets of models,
which are based on the TM1 and IUFSU parametrizations. In one set of models,
all saturation properties are the same except the symmetry energy slope $L$.
We determine the model parameters by adjusting simultaneously $g_{\rho}$ and
${\Lambda}_{\textrm{v}}$ so as to get a given $L$ and keep $E_{\textrm{sym}}$
unchanged at saturation density. For the TM1 case, we consider that $L$ varies
from $50$ MeV to $110.8$ MeV which is the value of the original TM1 model.
For the IUFSU case, the range of $L$ considered is relatively small,
$47.2 \leq L \leq 80$ MeV. This is because smaller $L$ like 40 MeV
or larger $L$ like 90 MeV in the IUFSU case will predict negative pressures
at some densities in pure neutron matter, which is not suitable for
the description of dripped neutron gas existing in the inner crust
of neutron stars. Using the set of models with different $L$,
it is possible to study the impact of $L$
on the crust-core transition and pasta phase properties.
In Ref.~\cite{Mene11}, the authors have kept the symmetry energy fixed at
the density 0.12 fm$^{-3}$. We prefer to fix $E_{\textrm{sym}}$ and vary $L$
at saturation density so that it is easy to compare
with other studies~\cite{Oyam07,LiBA08,Gril12}.
In the present work, we use two different RMF models (TM1 and IUFSU)
to examine the model dependence of the results obtained.
In Tables~\ref{tab:2} and~\ref{tab:3}, we present the parameters,
$g_{\rho}$ and ${\Lambda}_{\textrm{v}}$,
generated from the TM1 and IUFSU models for different symmetry energy
slope $L$ and fixed symmetry energy $E_{\textrm{sym}}$ at saturation density.

We adopt the Wigner-Seitz approximation to describe the inner crust matter,
which consists of nuclei surrounded by electron and neutron gases.
The conditions of $\beta$ equilibrium and charge neutrality are assumed
to be satisfied inside the Wigner-Seitz cell.
We use the coexisting phases method~\cite{Mene08,Maru05,SSA10,SSA09}
to study the properties of the Wigner-Seitz cell.
The matter inside the cell separates into two coexisting
phases with a sharp interface. The dense and dilute phases are denoted
by phase \Rmnum{1} and phase \Rmnum{2}, respectively.
These two phases satisfy Gibbs conditions for phase equilibrium.
In order to obtain the properties of phase \Rmnum{1} and phase \Rmnum{2},
we simultaneously solve the following equations:
\begin{eqnarray}
\label{eq:CP1}
 & & P^{\textrm{\Rmnum{1}}} = P^{\textrm{\Rmnum{2}}},  \\
\label{eq:CP2}
 & & {\mu}^{\textrm{\Rmnum{1}}}_{i} = {\mu}^{\textrm{\Rmnum{2}}}_{i}, \hspace{1cm} i = p,n,  \\
\label{eq:CP3}
 & & m^2_{\sigma}{\sigma}^{\textrm{\Rmnum{1}}}+g_2\left({{\sigma}^{\textrm{\Rmnum{1}}}}\right)^2
+g_3\left({{\sigma}^{\textrm{\Rmnum{1}}}}\right)^3
= -g_{\sigma}{n}^{\textrm{\Rmnum{1}}}_{s},   \\
\label{eq:CP4}
 & & m^2_{\sigma}{\sigma}^{\textrm{\Rmnum{2}}}+g_2\left({{\sigma}^{\textrm{\Rmnum{2}}}}\right)^2
+g_3\left({{\sigma}^{\textrm{\Rmnum{2}}}}\right)^3
= -g_{\sigma}{n}^{\textrm{\Rmnum{2}}}_{s},   \\
\label{eq:CP5}
 & & m^2_{\omega}{\omega}^{\textrm{\Rmnum{1}}}+{c_3}\left({{\omega}^{\textrm{\Rmnum{1}}}}\right)^3
+2\Lambda_{\textrm{v}}g^2_{\omega}{\omega}^{\textrm{\Rmnum{1}}}
g^2_{\rho}\left({{\rho}^{\textrm{\Rmnum{1}}}}\right)^2
= g_{\omega}{n}^{\textrm{\Rmnum{1}}}_b,  \\
\label{eq:CP6}
 & & m^2_{\omega}{\omega}^{\textrm{\Rmnum{2}}}+{c_3}\left({{\omega}^{\textrm{\Rmnum{2}}}}\right)^3
+2\Lambda_{\textrm{v}}g^2_{\omega}{\omega}^{\textrm{\Rmnum{2}}}
g^2_{\rho}\left({{\rho}^{\textrm{\Rmnum{2}}}}\right)^2
= g_{\omega}{n}^{\textrm{\Rmnum{2}}}_b,
\end{eqnarray}
where $n_s^{\textrm{\Rmnum{1}}(\textrm{\Rmnum{2}})}$ and
      $n_b^{\textrm{\Rmnum{1}}(\textrm{\Rmnum{2}})}$ are
the scalar and vector densities of baryons in phases \Rmnum{1} and \Rmnum{2}, respectively.
We assume a uniform distribution for electrons inside the Wigner-Seitz cell.
The chemical potential of electrons is determined by the $\beta$ equilibrium condition,
$\mu_e=\mu_n^{\textrm{\Rmnum{1}}}-\mu_p^{\textrm{\Rmnum{1}}}$.
Furthermore, the Fermi momentum and number density of electrons are obtained by
$\sqrt{\left(k^e_F\right)^2 + m_e^2}=\mu_e$ and
$n_e = \left(k^e_F\right)^3 / 3\pi^2$.
The total pressure $P$ including contributions from baryons and electrons is given by
$P=P_b+P_e$ with $P_b=P^{\textrm{\Rmnum{1}}}=P^{\textrm{\Rmnum{2}}}$.
We consider a charge neutral cell, where the electron density $n_e$ is equal to
the average proton density $n_p$. The volume fraction $f$ of phase \Rmnum{1}
is determined by
\begin{equation}
 n_e=n_p=f n_p^{\textrm{\Rmnum{1}}}+ \left(1-f\right) n_p^{\textrm{\Rmnum{2}}}.
\end{equation}
The total energy density of the system is given by
\begin{equation}
\label{eq:ews}
\varepsilon=f{\varepsilon}^{\textrm{\Rmnum{1}}}+\left({1-f}\right)
    {\varepsilon}^{\textrm{\Rmnum{2}}}+{\varepsilon}_e+
    {\varepsilon}_{\textrm{surf}}+{\varepsilon}_{\textrm{Coul}},
\end{equation}
where ${\varepsilon}_e$, ${\varepsilon}_{\textrm{surf}}$,
and ${\varepsilon}_{\textrm{Coul}}$ denote the electron, surface, and
Coulomb energy densities, respectively.
It is known that nuclear pasta phase is mainly determined by
the competition between the surface and Coulomb energies~\cite{Wata00}.
The surface energy density is expressed as
\begin{equation}
{\varepsilon}_{\textrm{surf}}=\frac{{\tau}FD}{R_D},
\label{eq:es}
\end{equation}
where $\tau$ is the surface tension,
$D=1,2,3$ is the geometrical dimension of the system,
and $R_D$ is the radius of the droplet (rod or slab).
$F$ is the volume fraction of the inner part, we take $F=f$
for droplets, rods, and slabs, while $F =1-f$ for bubbles and tubes.
The Coulomb energy density is given by
\begin{equation}
{\varepsilon}_{\textrm{Coul}}=
2{\pi}{e^2}\left(n^{\textrm{\Rmnum{1}}}_p-n^{\textrm{\Rmnum{2}}}_p\right)^2R^2_D F\Phi,
\end{equation}
where
\begin{equation}
\Phi=\left\{
\begin{array}{ll}
\frac{1}{D+2}\left(\frac{2-DF^{1-2/D}}{D-2}+F\right),  & D=1,3, \\
\frac{F-1-\ln{F}}{D+2},  & D=2. \\
\end{array} \right.
\end{equation}
By minimizing ${\varepsilon}_{\textrm{surf}}+{\varepsilon}_{\textrm{Coul}}$ with respect
to $R_D$, we get ${\varepsilon}_{\textrm{surf}}=2{\varepsilon}_{\textrm{Coul}}$.
The radii of the droplet (rod, slab) and that of the Wigner-Seitz cell are respectively
given by
\begin{eqnarray}
\label{eq:RD}
R_D &=& \left[\frac{\tau{D}}{4{\pi}e^2
\left(n^{\textrm{\Rmnum{1}}}_p-n^{\textrm{\Rmnum{2}}}_p\right)^2\Phi}\right]^{1/3}, \\
\label{eq:RW}
R_W &=& \frac{R_D}{F^{1/D}}.
\end{eqnarray}

The surface tension $\tau$ plays a crucial role in determining
the crust-core transition in neutron stars~\cite{Mene10,Maru05}.
In Ref.~\cite{Maru05}, the authors have shown that the appearance of
pasta phases essentially depends on the value of the surface tension.
It has been found in Ref.~\cite{Mene08} that a parametrized surface tension
may fail to predict the appearance of the pasta phase in $\beta$ equilibrium matter.
Therefore, it is very important to determine the surface tension
in a proper manner. In this work, we calculate the surface
tension using a Thomas-Fermi approach with the same RMF
parametrization as that used in the coexisting phases.
We adopt the method described in~\cite{Mene10,Cent98,Douc00},
where the surface tension is calculated by a one-dimensional
system consisting of protons and neutrons.
We consider a semi-infinite slab with a plane interface which
separates a dense matter from a dilute neutron-rich matter.
The axis perpendicular to the interface is taken to be the $z$ axis.
When $z$ goes to $-\infty$ ($+\infty$), the neutron and proton
densities approach the values of phase \textrm{\Rmnum{1}} (\textrm{\Rmnum{2}}),
which are achieved by solving Eqs.~(\ref{eq:CP1})-(\ref{eq:CP6}).
With the density profiles obtained in the Thomas-Fermi approach,
we calculate the surface tension as~\cite{Cent98}
\begin{equation}
\tau=\int_{-\infty}^{\infty}dz\left\{\varepsilon\left(z\right)-\varepsilon^{\textrm{\Rmnum{2}}}-
\mu_p\left[n_p\left(z\right)-n^{\textrm{\Rmnum{2}}}_{p}\right]-
\mu_n\left[n_n\left(z\right)-n^{\textrm{\Rmnum{2}}}_{n}\right]\right\}.
\label{eq:sigma1}
\end{equation}
Also, the surface tension can be obtained from the derivatives of
the meson fields as~\cite{Mene10}
\begin{equation}
\tau=\int_{-\infty}^{\infty}dz\left[\left(\frac{d\sigma}{dz}\right)^2-
\left(\frac{d\omega}{dz}\right)^2-\left(\frac{d\rho}{dz}\right)^2\right].
\label{eq:sigm2}
\end{equation}
In Ref.~\cite{Mene10}, the authors have checked numerically the
equivalence between Eqs.~(\ref{eq:sigma1}) and (\ref{eq:sigm2}).
Here, we confirm that the values of the surface tension calculated by
Eqs.~(\ref{eq:sigma1}) and (\ref{eq:sigm2}) are very close with
each other.

We calculate the energy density of the system by Eq.~(\ref{eq:ews})
at a given density $n_b$ for all nuclear shapes considered
(droplet, rod, slab, tube, and bubble). The stable shape of the pasta phase
is finally taken as the one with the lowest energy density
for zero temperature matter.
The energy density of corresponding homogeneous phase at the same $n_b$
is also calculated and compared with that of the pasta phase.
The crust-core transition occurs at the density where
the homogeneous phase has a lower energy density than the pasta phase.
Using modified versions of TM1 (IUFSU) with different $L$,
it is possible to study the influence of $L$ on the crust-core transition
and pasta phase properties.

\section{Results and discussion}
\label{sec:3}

\begin{figure}[htb]
\includegraphics[bb=40 94 568 759, width=7 cm,clip]{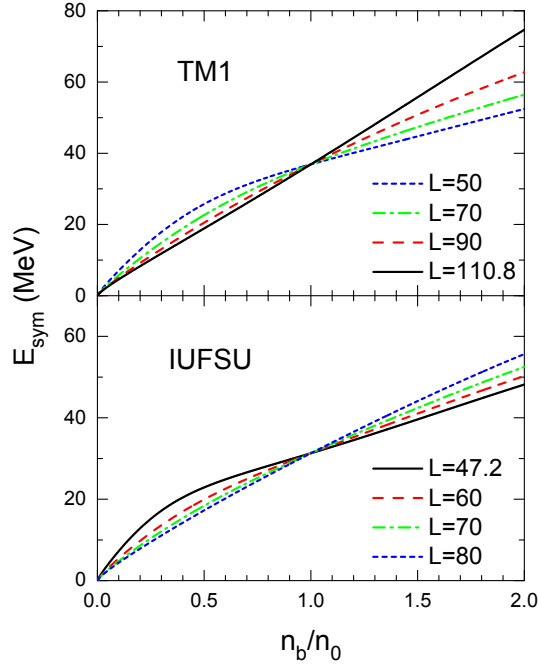}
\caption{(Color online) Symmetry energy, $E_{\textrm{sym}}$,
as a function of the ratio of baryon density to saturation density, $n_b/n_0$,
for modified versions of TM1 (upper panel) and IUFSU (lower panel)
with different $L$. The symmetry energy at saturation density is fixed at
$E_{\textrm{sym}}=36.9\, (31.3)$ MeV for TM1 (IUFSU).}
\label{fig:1Esym}
\end{figure}

In this section, we investigate the effect of the symmetry energy slope $L$
on properties of pasta phases and the crust-core transition.
In order to study the influence of $L$ clearly, we use two sets of models generated
from the TM1 and IUFSU parametrizations. Note that all models in each set have the same
symmetry energy $E_{\textrm{sym}}$ but different slope $L$ at saturation density.
In Fig.~\ref{fig:1Esym}, we plot the symmetry energy, $E_{\textrm{sym}}$,
as a function of the ratio of baryon density to saturation density, $n_b/n_0$,
for the set of models generated from TM1 (upper panel) and IUFSU (lower panel).
The symmetry energy is fixed at the original value $36.9$ (31.3) MeV
at saturation density in the case of TM1 (IUFSU). It is shown that a larger $L$
corresponds to a smaller $E_{\textrm{sym}}$ at subnuclear densities, and, as a result,
favors a more neutron-rich matter for homogeneous phase in comparison with a smaller $L$.
The density dependence of $E_{\textrm{sym}}$ would be the main reason for
correlations between $L$ and the crust-core transition and pasta phase properties.
On the other hand, the density dependence of $E_{\textrm{sym}}$ also affects
the masses and radii of neutron stars~\cite{Mene11,Prov13}.
In general, the star radius increases with the symmetry energy slope $L$,
while the maximum mass does not strongly depend on $L$.
We calculate neutron-star masses using the two sets of models
generated from the TM1 and IUFSU parametrizations in order to
test their compatibility with the largest well measured mass
$\sim 1.97 \pm 0.04 \ M_\odot$ of PSR J1614-2230~\cite{Demo10}.
In the case of TM1, the maximum mass of the original model ($L=110.8$ MeV)
is about $2.18\ M_\odot$, while it decreases to $2.11\ M_\odot$ for $L=50$ MeV.
In the case of IUFSU, the maximum mass of the original model ($L=47.2$ MeV)
is about $1.94\ M_\odot$, while it increases to $1.95\ M_\odot$ for $L=80$ MeV.
We confirm that all models used in the present work are compatible
with the mass of PSR J1614-2230.

\begin{figure}[htb]
\includegraphics[bb=32 87 557 750, width=7.0 cm,clip]{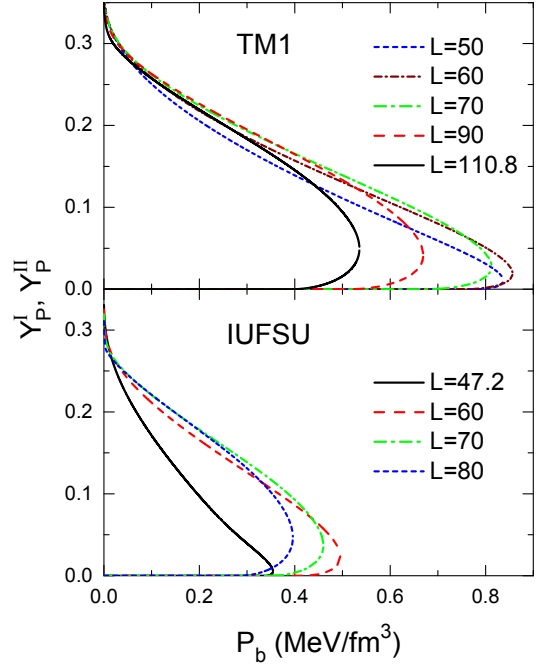}
\caption{(Color online) Proton fractions of dense phase (phase \textrm{\Rmnum{1}})
and dilute phase (phase \textrm{\Rmnum{2}}), $Y_p^{\textrm{\Rmnum{1}}}$ and
$Y_p^{\textrm{\Rmnum{2}}}$, as a function of the baryon pressure $P_b$
for modified versions of TM1 (upper panel) and IUFSU (lower panel)
with different $L$. The pairs of solutions form the boundary of
the coexistence phases (binodal curve).}
\label{fig:2CP-Yp}
\end{figure}

\begin{figure}[htb]
\includegraphics[bb=27 87 557 750, width=7.0 cm,clip]{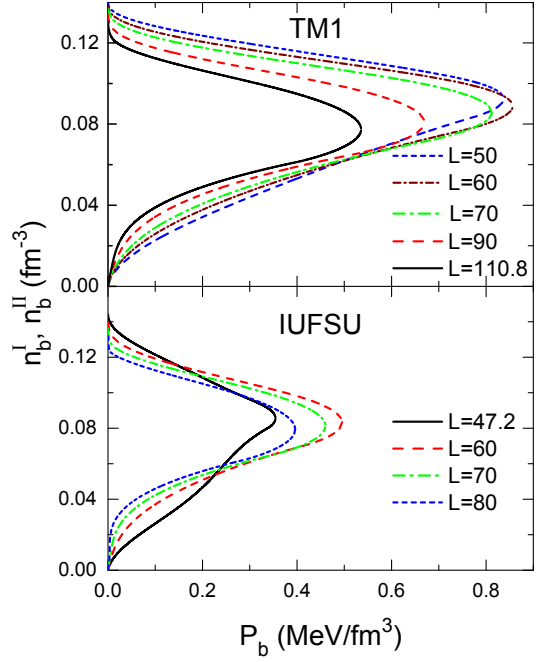}
\caption{(Color online) Baryon densities of dense phase (phase \textrm{\Rmnum{1}})
and dilute phase (phase \textrm{\Rmnum{2}}), $n_b^{\textrm{\Rmnum{1}}}$ and
$n_b^{\textrm{\Rmnum{2}}}$, as a function of the baryon pressure $P_b$
for modified versions of TM1 (upper panel) and IUFSU (lower panel)
with different $L$.}
\label{fig:3CP-nb}
\end{figure}

In the present study, we focus on the correlation between the symmetry energy
slope $L$ and the crust-core transition.
We consider the inner crust matter consisting of protons, neutrons, and electrons
in $\beta$ equilibrium. The coexisting phases method is used to describe
the matter inside the Wigner-Seitz cell, in which the nuclear matter
is assumed to separate into two coexisting phases:
a dense phase (phase \textrm{\Rmnum{1}}) and a dilute neutron-rich
phase (phase \textrm{\Rmnum{2}}).
The electrons are treated as a uniform gas in the cell,
which does not affect the two coexisting phases of nuclear matter.
We obtain the properties of phase \textrm{\Rmnum{1}} and phase \textrm{\Rmnum{2}}
by solving the Gibbs conditions, Eqs.~(\ref{eq:CP1})-(\ref{eq:CP6}),
at a given baryon pressure $P_b$.
In Fig.~\ref{fig:2CP-Yp}, we plot the proton fractions, $Y_p^{\textrm{\Rmnum{1}}}$ and
$Y_p^{\textrm{\Rmnum{2}}}$, as functions of $P_b$.
The pairs of solutions, $Y_p^{\textrm{\Rmnum{1}}}$ and $Y_p^{\textrm{\Rmnum{2}}}$,
form the boundary of the coexistence phases (binodal curve).
At lower pressure, we obtain a large $Y_p^{\textrm{\Rmnum{1}}}$ together with
$Y_p^{\textrm{\Rmnum{2}}}=0$, which corresponds to the case where there is
a pure neutron gas coexisting with dense nuclear matter.
As pressure increases, $Y_p^{\textrm{\Rmnum{1}}}$ decreases and protons
begin to drip at $Y_p^{\textrm{\Rmnum{2}}}>0$. For the original
TM1 (IUFSU) model with $L=110.8\, (47.2)$ MeV, protons begin to drip
at $P_b=0.40 \, (0.34) \, \textrm{MeV/fm}^3$.
It is shown that the proton drip point increases with decreasing $L$,
but it begins to decrease for $L<60$ MeV.
Moreover, as the pressure increases the system encounters a critical
pressure beyond which the two coexisting phases disappear.
This critical pressure is related to nuclear liquid-gas phase transition
without Coulomb and surface effects.
It is clearly seen that the critical pressure depends on the symmetry energy slop $L$.
For the large-$L$ region ($L>60$ MeV), the critical pressure
increases with decreasing $L$ in both TM1 and IUFSU cases
as shown in Fig.~\ref{fig:2CP-Yp}.
We present in Fig.~\ref{fig:3CP-nb} the pair of baryon densities,
$n_b^{\textrm{\Rmnum{1}}}$ and $n_b^{\textrm{\Rmnum{2}}}$, as a function of $P_b$.
At lower pressure, $n_b^{\textrm{\Rmnum{1}}}$ is close to normal nuclear matter
density, while the density of dripped neutrons $n_b^{\textrm{\Rmnum{2}}}$ is very small.
With increasing pressure, $n_b^{\textrm{\Rmnum{1}}}$ and $n_b^{\textrm{\Rmnum{2}}}$
get close to each other. At the critical point, the two coexisting phases
disappear. It is found that the density at the critical point
increases with decreasing $L$ in both TM1 and IUFSU cases.
We can see in both Figs.~\ref{fig:2CP-Yp} and~\ref{fig:3CP-nb} that
there is a turnaround in the dependence of the critical pressure on
the symmetry energy slop $L$ at small $L$ ($L<60$ MeV).
This may be understood as a result of several competing effects.
First, the pressure of neutron-rich matter at fixed density and proton fraction
generally increases with increasing $L$~\cite{Duco11}. But second,
the density at the critical point decreases with increasing $L$, which causes
a decrease of the pressure. Third, the shift of the proton fraction
at the critical point may also affect the critical pressure.
These competing effects lead to a nontrivial dependence of
the critical pressure on $L$ as shown in Figs.~\ref{fig:2CP-Yp} and~\ref{fig:3CP-nb}.

\begin{figure}[htb]
\includegraphics[bb=18 85 568 759, width=7.0 cm,clip]{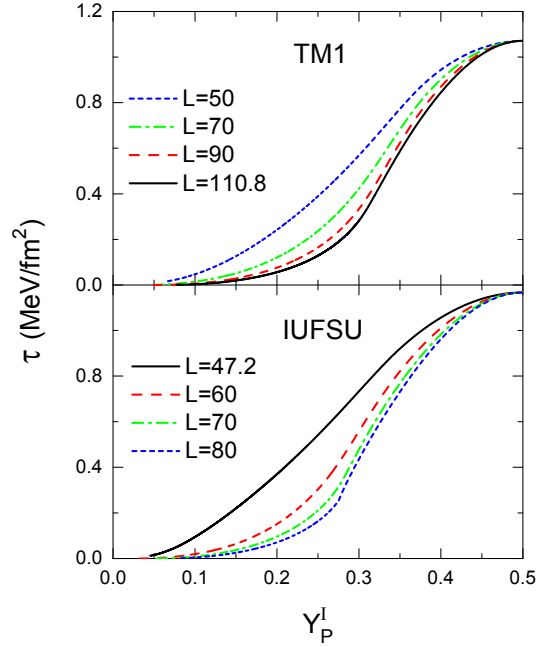}
\caption{(Color online) Surface tension $\tau$ as a function of
the proton fraction of the dense phase, $Y^{\textrm{\Rmnum{1}}}_P$,
for modified versions of TM1 (upper panel) and IUFSU (lower panel)
with different $L$.}
\label{fig:4CP-T}
\end{figure}

In Fig.~\ref{fig:4CP-T}, we show the surface tension $\tau$ as a function
of the proton fraction in the dense phase, $Y_p^{\textrm{\Rmnum{1}}}$,
which is obtained using a Thomas-Fermi approach. The surface tension $\tau$ is calculated
by a one-dimensional system consisting of protons and neutrons.
The densities at minus (plus) infinity are close to the values of phase \textrm{\Rmnum{1}}
(\textrm{\Rmnum{2}}), while the density profile is obtained in the Thomas-Fermi approach.
It is shown that $\tau$ in all cases decreases monotonically with decreasing
$Y_p^{\textrm{\Rmnum{1}}}$. As for the dependence of $\tau$ on $L$,
it is found that $\tau$ for all $L$ is almost identical near
$Y_p^{\textrm{\Rmnum{1}}}=0.5$ due to the same isoscalar properties among
models in each set. As $Y_p^{\textrm{\Rmnum{1}}}$ decreases, $\tau$ shows
a clear dependence on $L$. A smaller $L$ produces a larger $\tau$, which is
consistent with other studies~\cite{Gril12,SSA10}.
In the set of IUFSU, the original IUFSU model has the smallest $L$ ($47.2$ MeV),
so it has the largest surface tension. For the case of TM1, the original
TM1 model has the largest $L$ ($110.8$ MeV), so it has the smallest $\tau$.
Considering the crucial role of $\tau$ in determining the crust-core
transition, we do not take any parametrized form for $\tau$ in the present study,
but prefer to use the exact values obtained in the Thomas-Fermi approach.
We have checked that using a parametrized form given in Ref.~\cite{SSA10}
for $\tau$ may cause a slight difference for the crust-core transition density
in comparison with using the exact values.
For instance, the crust-core transition density obtained using the
parametrized form given by Eq. (41) of Ref.~\cite{SSA10} for the NL3
parametrization is 0.0496 $\textrm{fm}^{-3}$, while it is 0.0522 $\textrm{fm}^{-3}$
with the surface tension using the exact values calculated by Eq.~(\ref{eq:sigm2}).
We confirm that the proton fraction in Eq. (41) of Ref.~\cite{SSA10}
should be the one in the dense phase, $Y_p^{\textrm{\Rmnum{1}}}$ (see the text
below Eq. (41) of Ref.~\cite{SSA10}). Therefore, the crust-core transition
density (0.068 $\textrm{fm}^{-3}$) given in the Erratum of Ref.~\cite{Mene10}
(see the sixth line in Table 1) seems to be questionable
since the global proton fraction
has been used for Eq. (41) of Ref.~\cite{SSA10} in their calculation.
However, if another parametrized form defined by Eq. (45) of Ref.~\cite{Mene08}
is used for the same case, the homogeneous phase always has the lowest energy density,
and as a result the existence of pasta phases in neutron star crusts could not
be achieved. The failure in predicting the existence of pasta phases is due to
Eq. (45) giving too large surface tension as shown
in Fig. 6(c) of Ref.~\cite{Mene08}.
By comparing results with different treatments of $\tau$,
we find that the surface tension plays a very important role in determining
the crust-core transition.
On the other hand, different treatments of $\tau$ will not change the onset densities of
various shapes in pasta phases, which is because the stable shape
is determined by the energy difference between various shapes.
We note that only the sum
${\varepsilon}_{\textrm{surf}}+{\varepsilon}_{\textrm{Coul}}$
in Eq.~(\ref{eq:ews}) depends on $\tau$, which is proportional
to $\tau^{2/3}$ based on Eqs.~(\ref{eq:es}) and~(\ref{eq:RD}).
The transition between two pasta shapes occurs at the density where
their energy difference changes sign, but this cannot be altered
by the value of $\tau$.
In the present study, we perform a self-consistent calculation
for $\tau$ within the Thomas-Fermi approach.

\begin{figure}[htb]
\includegraphics[bb=61 94 557 750, width=7.0 cm,clip]{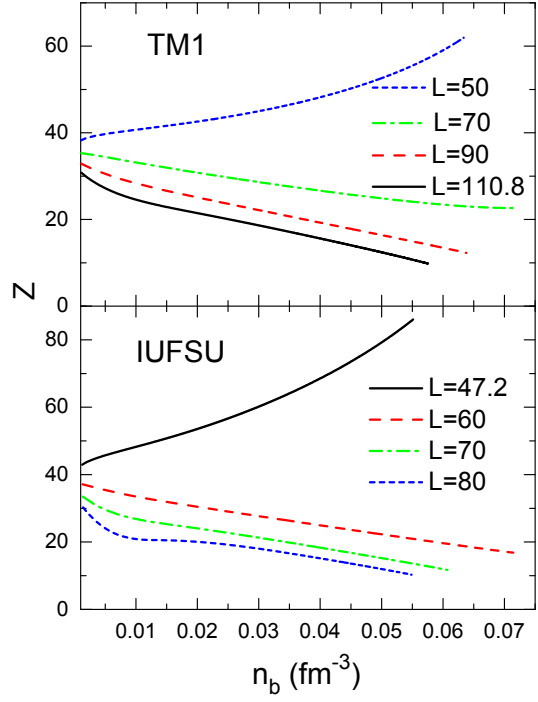}
\caption{(Color online) Proton number $Z$ inside spherical nuclei
as a function of the baryon density $n_b$, for modified versions
of TM1 (upper panel) and IUFSU (lower panel) with different $L$.}
\label{fig:5Z}
\end{figure}

\begin{figure}[htb]
\includegraphics[bb=35 88 557 750, width=7.0 cm,clip]{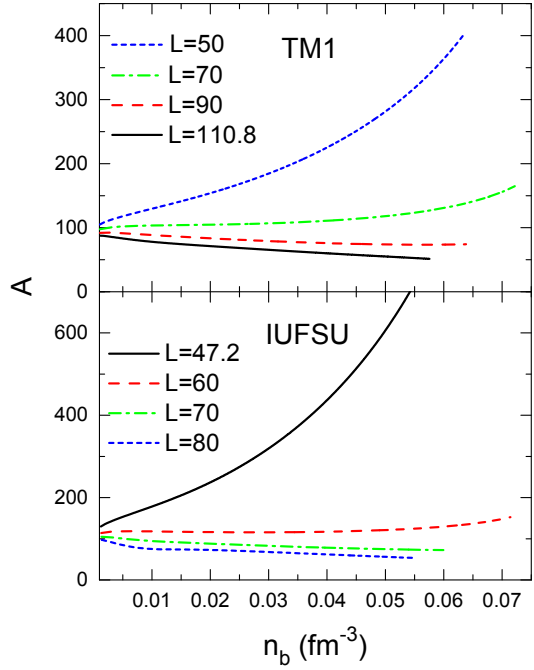}
\caption{(Color online) Same as Fig.~\ref{fig:5Z}, but for
the nucleon number $A$ inside spherical nuclei.}
\label{fig:6A}
\end{figure}

\begin{figure}[htb]
\includegraphics[bb=41 89 556 750, width=7.0 cm,clip]{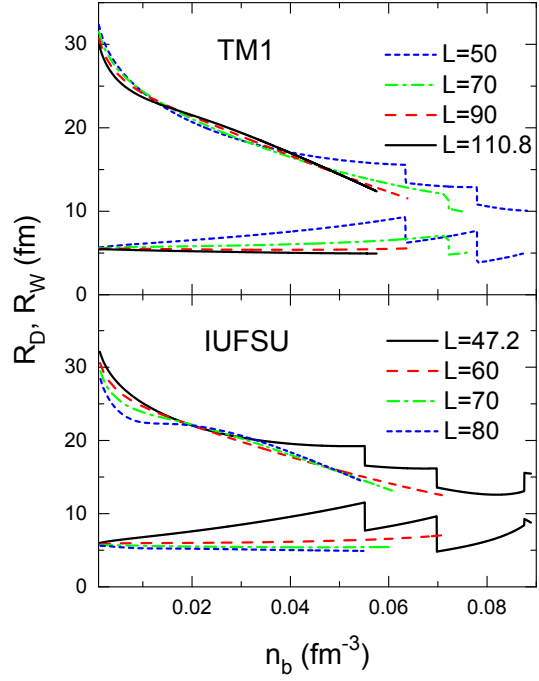}
\caption{(Color online) Nuclear radius $R_D$ and Wigner-Seitz
cell radius $R_W$ as a function of the baryon density $n_b$
for modified versions of TM1 (upper panel) and IUFSU (lower panel)
with different $L$. The jumps in $R_D$ and $R_W$ at $n_b>0.05$ fm$^{-3}$
correspond to shape transitions in pasta phases.}
\label{fig:7R}
\end{figure}

\begin{figure}[htb]
\includegraphics[bb=16 478 573 729, width=12.0 cm,clip]{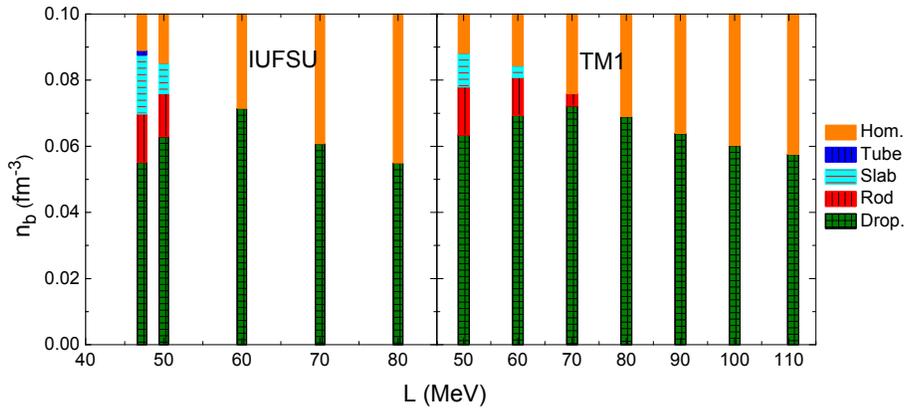}
\caption{(Color online) Phase diagrams for modified versions of
TM1 (right panel) and IUFSU (left panel).
Different colors represent droplet, rod, slab, tube, and
homogeneous phases as given in the legend.}
\label{fig:8PP}
\end{figure}

For pasta phases, we consider five nuclear shapes: droplet,
rod, slab, tube, and bubble. The most stable shape among them is the one
with the lowest energy density calculated by Eq.~(\ref{eq:ews}).
At lower densities, the shape of stable nuclei is spherical (droplet).
In Figs.~\ref{fig:5Z} and \ref{fig:6A}, we plot the proton number $Z$
and nucleon number $A$ inside spherical nuclei as a function of
the average baryon density $n_b$. For larger $L$, it is found that
both $Z$ and $A$ decrease with increasing density, but opposite behavior
is observed for smaller $L$, as in the case of the original IUFSU model.
The density dependence of $Z$ and $A$ is consistent with that obtained
in other works~\cite{Oyam07,Gril12}.
On the other hand, $Z$ and $A$ increase monotonically with decreasing $L$
at a fixed density $n_b$. This $L$ dependence is related to the behavior
of the surface tension $\tau$ as shown in Fig.~\ref{fig:4CP-T}.
It could be understood from the size equilibrium condition,
${\varepsilon}_{\textrm{surf}}=2{\varepsilon}_{\textrm{Coul}}$,
which gives the result that the proton number $Z$ increases with increasing surface
tension $\tau$ and nuclear radius $R_D$. Therefore, a smaller $L$
favors larger $\tau$, $Z$, $A$, and $R_D$.
In Fig.~\ref{fig:7R}, we show the resulting radius of the nucleus $R_D$
and that of the Wigner-Seitz cell $R_W$ given by Eqs.~(\ref{eq:RD})
and (\ref{eq:RW}). It is found that $R_W$ decreases with density clearly,
whereas $R_D$ does not have obvious density dependence for large $L$
at low densities. This tendency is related to the increase of nuclear
volume fraction with density. The jumps in $R_D$ and $R_W$ at $n_b>0.05$ fm$^{-3}$
correspond to shape transitions in pasta phases.
In Fig.~\ref{fig:8PP}, we display the density range of various pasta phases
for modified versions of TM1 (right panel) and IUFSU (left panel)
with different $L$, while the transition to the homogeneous phase
is also presented. It is seen that only droplet configuration
appears before the crust-core transition for larger $L$,
namely $L\geq 80$ (60) MeV in the case of TM1 (IUFSU).
As $L$ decreases, other pasta shapes may occur in the sequence of
rod, slab, and tube configurations as shown in Fig.~\ref{fig:8PP},
but the bubble configuration does not appear in all cases considered.
In the original IUFSU model ($L=47.2$ MeV), the onset densities of
various pasta phases are close to those
given in Ref.~\cite{Gril12} using the Thomas-Fermi method.
Comparing the behavior of TM1 with that of IUFSU, we find that
the two sets of models have a similar $L$ dependence.
These results also agree with those obtained in Refs.~\cite{Oyam07,Gril12}.

\begin{figure}[htb]
\includegraphics[bb=27 221 555 710, width=7.0 cm,clip]{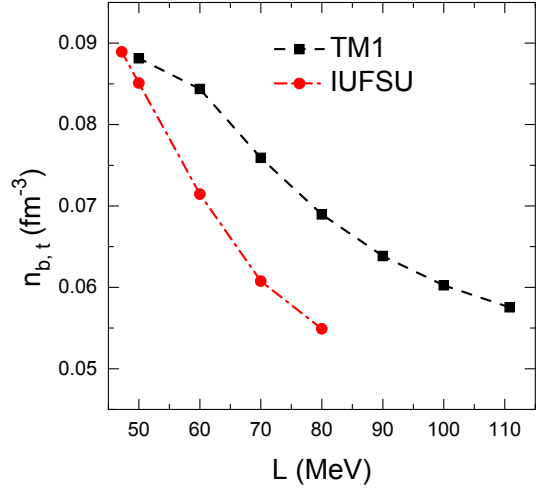}
\caption{(Color online) Baryon density at the crust-core transition $n_{b,t}$
as a function of the symmetry energy slope $L$, obtained for
the set of models generated from TM1 and IUFSU parametrizations.}
\label{fig:9TR}
\end{figure}

\begin{figure}[htb]
\includegraphics[bb=33 221 555 710, width=7.0 cm,clip]{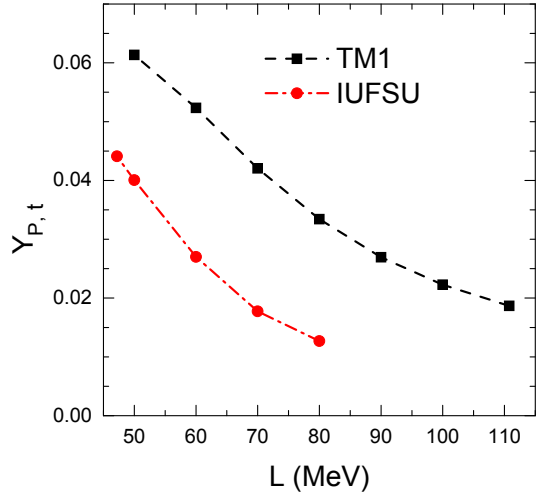}
\caption{(Color online) Proton fraction at the crust-core transition
$Y_{P,t}$ as a function of the symmetry energy slope $L$, obtained for
the set of models generated from TM1 and IUFSU parametrizations.}
\label{fig:10TYP}
\end{figure}

\begin{figure}[htb]
\includegraphics[bb=38 221 556 709, width=7.0 cm,clip]{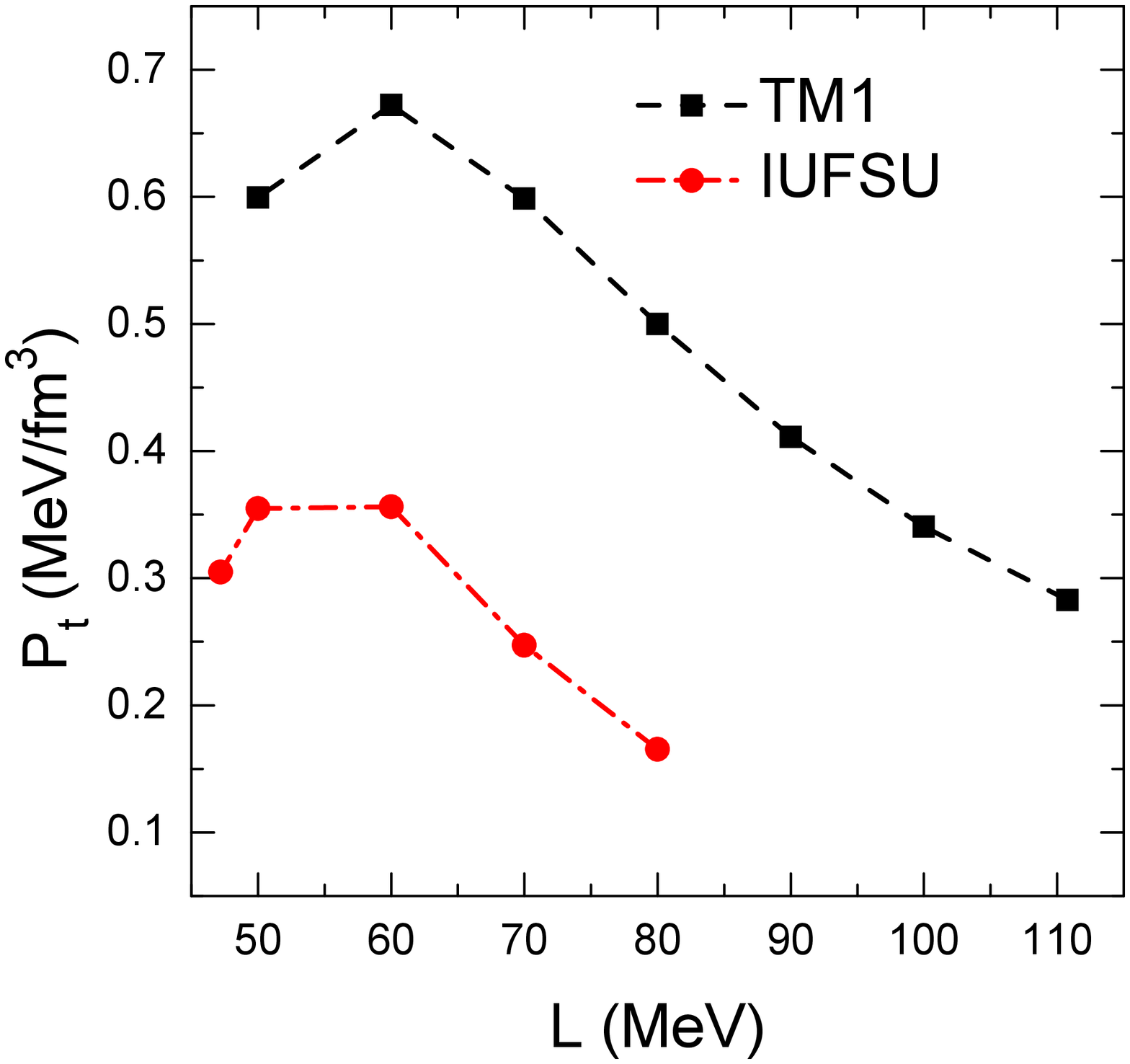}
\caption{(Color online) Pressure at the crust-core transition
$P_t$ as a function of the symmetry energy slope $L$, obtained for
the set of models generated from TM1 and IUFSU parametrizations.}
\label{fig:11TP}
\end{figure}

The crust-core transition occurs at the density where the homogeneous phase
has a lower energy density than the pasta phase.
It is important to investigate possible correlations between $L$ and
the crust-core transition.
In Fig.~\ref{fig:9TR}, we plot the crust-core transition density $n_{b,t}$
as a function of $L$. It is seen that there exists a clear correlation
between $L$ and $n_{b,t}$. This result is in good agreement with
those obtained by other methods~\cite{Oyam07,Duco11,Horo01,Horo03}.
In the case of the original TM1 model, we obtain $n_{b,t}=0.058 \, \textrm{fm}^{-3}$,
while the crust-core transition densities obtained from the dynamical and
thermodynamical methods are respectively 0.06 and 0.07 $\textrm{fm}^{-3}$ as
given in Table \textrm{\Rmnum{2}} of Ref.~\cite{Duco11}.
The difference is mainly due to the Coulomb and surface effects
being taken into account in the coexisting phases method used here,
while they are not included in the thermodynamical method.
In Fig.~\ref{fig:10TYP}, we display the proton fraction at
the crust-core transition $Y_{p,t}$ as a function of $L$.
It is clear that $Y_{p,t}$ decreases
with increasing $L$. These results are consistent with those obtained
in Ref.~\cite{Duco11}.
The tendency is due to a larger $L$ being related to a smaller symmetry
energy at subnuclear densities, and, as a result, corresponding to a smaller
proton fraction in $\beta$ equilibrium matter.
In Fig.~\ref{fig:11TP}, the core-crust transition pressure $P_t$ is shown
as a function of $L$. We find that $P_t$ decreases with increasing $L$ for
large $L$ region ($L>60$ MeV) in both TM1 and IUFSU cases.
This observation is different from that obtained in Ref.~\cite{Duco11},
which concluded that no satisfactory correlation could be seen between
the transition pressure and $L$, which is very sensitive to the model used.
As shown in Fig.~\ref{fig:11TP}, the correlation between $L$ and $P_t$
becomes weaker close to $L \sim 60$ MeV, and this trend has also been observed
in Ref.~\cite{Duco11}.
The nontrivial dependence of $P_t$ on $L$ is again a result of
the competing effects discussed above.
For neutron-rich matter at fixed density and proton fraction,
the pressure should increase with increasing $L$.
However, the decrease of $n_{b,t}$ with $L$ as shown in Fig.~\ref{fig:9TR}
causes a decrease of $P_t$ with increasing $L$.
On the other hand, the decrease of $Y_{p,t}$ with $L$
is also expected to affect the dependence
of $P_t$ on $L$, but this effect is found to be very small~\cite{Duco10}.
The competing contributions to the variation of $P_t$ with $L$ have been
thoroughly analyzed using a generalized liquid-drop model in Ref.~\cite{Duco10}.
Here, we find that $P_t$ depends on $L$ nonmonotonically as shown
in Fig.~\ref{fig:11TP}, which is due to the balance between
the competing effects.

In closing this section, we discuss how sensitive the results obtained
depend on the RMF models used. By comparing the results of TM1 with those
of IUFSU, we find that they exhibit qualitatively similar behavior
in the crust-core transition and pasta phase properties, especially their
dependence on the symmetry energy slope $L$.
In both TM1 and IUFSU models, a larger $L$ favors a lower transition
density $n_{b,t}$, a smaller proton fraction $Y_{p,t}$, and a lower
transition pressure $P_t$ as shown in Figs.~\ref{fig:9TR},~\ref{fig:10TYP},
and~\ref{fig:11TP}. On the other hand, a smaller $L$ is related to a
larger surface tension $\tau$ and more complex pasta phases.
In general, the results obtained in this study are consistent
with those given by other methods. We confirm that there exist clear
correlations between the symmetry energy slope $L$ and the crust-core
transition and pasta phase properties.

\section{Conclusion}
\label{sec:4}

In this article, we have studied the effects of the symmetry energy
on nuclear pasta phases in neutron stars. Especially, we have
systematically examined possible correlations between the symmetry energy
slope $L$ and the crust-core transition and pasta phase properties.
We have employed the RMF theory and the coexisting phases method
to investigate the properties of pasta phases and the crust-core transition.
To study the influence of $L$, we have generated two sets of the RMF models
based on the TM1 and IUFSU parametrizations.
In one set of models, all saturation properties are the same
except the symmetry energy slope $L$ which is controlled by tuning
the $\omega$-$\rho$ coupling strength.
By using the set of models with different $L$, it is possible to
study the impact of $L$ on the crust-core transition and pasta phase
properties. The results have been compared with those obtained from
the dynamical and thermodynamical methods~\cite{Duco11} and with that of
the Thomas-Fermi calculations~\cite{Oyam07,Gril12}.

We have investigated the properties of pasta phases considering five
nuclear shapes: droplet, rod, slab, tube, and bubble.
It has been found that only the droplet configuration
appears before the crust-core transition for larger $L$,
namely $L\geq 80$ (60) MeV in the case of TM1 (IUFSU).
As $L$ decreases, other pasta shapes may occur in the sequence of
rod, slab, and tube configurations, but the bubble configuration
does not appear in all cases considered.
We have also calculated the proton number $Z$ and nucleon number $A$
inside spherical nuclei at low densities for various values of $L$.
For larger $L$, both $Z$ and $A$ decrease with increasing density,
but opposite behavior is observed for smaller $L$.
On the other hand, $Z$ and $A$ increase monotonically with decreasing
$L$ at a fixed density. These results are consistent with those obtained
in other works~\cite{Oyam07,Gril12}.
Considering the crucial role of the surface tension in determining
the crust-core transition, we have performed a self-consistent calculation
for the surface tension $\tau$ within the Thomas-Fermi approach.
It has been found that a smaller
$L$ produces a larger $\tau$, and as a result, relates to larger $Z$
and $A$ in spherical nuclei.

We have observed clear correlations between the symmetry energy slope $L$
and the crust-core transition density $n_{b,t}$, proton fraction $Y_{p,t}$,
and pressure $P_t$. It has been shown that both $n_{b,t}$ and $Y_{p,t}$
decrease with increasing $L$, which can be related to the density dependence
of the symmetry energy at subnuclear densities. These results are consistent
with those obtained by other methods~\cite{Oyam07,Duco11}.
As for the transition pressure $P_t$, we have found that $P_t$ decreases
with increasing $L$ for the large-$L$ region ($L>60$ MeV) in both TM1
and IUFSU cases. This observation is different from that obtained in
Ref.~\cite{Duco11}.

We have compared the results obtained using two different RMF models,
namely the TM1 and IUFSU parametrizations. It has been shown that they
exhibit qualitatively similar behavior in the crust-core transition
and pasta phase properties, especially their dependence on the symmetry
energy slope $L$. In general, the results obtained in this study are
consistent with those given by other methods. We note that nuclear
shell and paring effects are neglected in the present study.
It would be very interesting to examine how these effects influence
various properties of pasta phases.

\section*{Acknowledgment}

This work was supported in part by the National Natural
Science Foundation of China (Grants No. 11075082 and No.
11375089).


\newpage
\begin{table}[tbp]
\caption{Parameter sets used in this work. The masses are given in MeV.}
\begin{center}
\begin{tabular}{lccccccccccc}
\hline\hline
Model   &$M$  &$m_{\sigma}$  &$m_\omega$  &$m_\rho$  &$g_\sigma$  &$g_\omega$
        &$g_\rho$ &$g_{2}$ (fm$^{-1}$) &$g_{3}$ &$c_{3}$ &$\Lambda_{\textrm{v}}$ \\
\hline
TM1     &938.0  &511.198  &783.0  &770.0  &10.0289  &12.6139  &9.2644
        &-7.2325   &0.6183   &71.3075   &0.000  \\
IUFSU   &939.0  &491.500  &782.5  &763.0  &9.9713   &13.0321  &13.5899
        &-8.4929   &0.4877   &144.2195  &0.046 \\
\hline\hline
\end{tabular}
\label{tab:1}
\end{center}
\end{table}

\begin{table}[tbp]
\caption{Parameters $g_{\rho}$ and ${\Lambda}_{\textrm{v}}$ generated
from the TM1 model for different symmetry energy slope $L$
and fixed symmetry energy $E_{\textrm{sym}}=36.9$ MeV at saturation density.
The original TM1 model has $L=110.8$ MeV.}
\begin{center}
\begin{tabular}{lccccccc}
\hline\hline
$L$ (MeV)               &50.0    &60.0     &70.0     &80.0     &90.0     &100.0   &110.8 \\
\hline
$g_{\rho}$              &13.8757 &12.6431  &11.6896  &10.9237  &10.2910  &9.7569  &9.2644\\
$\Lambda_{\textrm{v}}$  &0.0254  &0.0212   &0.0171   &0.0129   &0.0087   &0.0045  &0.0000\\
\hline\hline
\end{tabular}
\label{tab:2}
\end{center}
\end{table}

\begin{table}[tbp]
\caption{Parameters $g_{\rho}$ and ${\Lambda}_{\textrm{v}}$ generated
from the IUFSU model for different symmetry energy slope $L$
and fixed symmetry energy $E_{\textrm{sym}}=31.3$ MeV at saturation density.
The original IUFSU model has $L=47.2$ MeV.}
\begin{center}
\begin{tabular}{lccccc}
\hline\hline
$L$ (MeV)              &47.2     &50.0     &60.0     &70.0   &80.0   \\
\hline
$g_{\rho}$             &13.5899  &12.6766  &10.4742  &9.1260 &8.1926 \\
$\Lambda_{\textrm{v}}$ &0.0460   &0.0433   &0.0336   &0.0238 &0.0141 \\
\hline\hline
\end{tabular}
\label{tab:3}
\end{center}
\end{table}

\newpage

\newpage
\begin{thebibliography}{99}

\bibitem{Webe05} F. Weber, Prog. Part. Nucl. Phys. \textbf{54}, 193 (2005).

\bibitem{PR00} H. Heiselberg and M. Hjorth-Jensen, Phys. Rep. \textbf{328}, 237 (2000).

\bibitem{PR07} J. M. Lattimer and M. Prakash, Phys. Rep. \textbf{442}, 109 (2007).

\bibitem{Stei98} A. W. Steiner, Phys. Rev. C \textbf{77}, 035805 (2008).

\bibitem{Rave83} D. G. Ravenhall, C. J. Pethick, and J. R. Wilson,
Phys. Rev. Lett. \textbf{50}, 2066 (1983).

\bibitem{Wata00} G. Watanabe, K. Iida, and K. Sato, Nucl. Phys. \textbf{A676}, 455 (2000).

\bibitem{Oyam07} K. Oyamatsu and K. Iida, Phys. Rev. C \textbf{75}, 015801 (2007).

\bibitem{Mene08} S. S. Avancini, D. P. Menezes, M. D. Alloy, J. R. Marinelli,
M. M. W. Moraes, and C. Provid\^{e}ncia, Phys. Rev. C \textbf{78}, 015802 (2008).

\bibitem{Mene10} S. S. Avancini, S. Chiacchiera, D. P. Menezes, and C. Provid\^{e}ncia,
Phys. Rev. C \textbf{82}, 055807 (2010);
\textbf{85}, 059904(E) (2012).

\bibitem{Gril12} F. Grill, C. Provid\^{e}ncia, and S. S. Avancini,
Phys. Rev. C \textbf{85}, 055808 (2012).

\bibitem{Duco11} C. Ducoin, J. Margueron, C. Provid\^{e}ncia, and I. Vidana,
Phys. Rev. C \textbf{83}, 045810 (2011).

\bibitem{LiBA08} B. A. Li, L. W. Chen, and C. M. Ko, Phys. Rep. \textbf{464}, 113 (2008).

\bibitem{Chen13} Z. Zhang and L. W. Chen, Phys. Lett. \textbf{B726}, 234 (2013).

\bibitem{Mene11} R. Cavagnoli, D. P. Menezes, and C. Provid\^{e}ncia,
Phys. Rev. C \textbf{84}, 065810 (2011).

\bibitem{Sero86} B. D. Serot and J. D. Walecka, Adv. Nucl. Phys.
\textbf{16}, 1 (1986).

\bibitem{Ring90} Y. K. Gambhir, P. Ring, and A. Thimet, Ann. Phys. (N.Y.)
\textbf{198}, 132 (1990).

\bibitem{Meng06} J. Meng, H. Toki, S. G. Zhou, S. Q. Zhang, W. H. Long, and L. S. Geng,
Prog. Part. Nucl. Phys. \textbf{57}, 470 (2006).

\bibitem{TM1} Y. Sugahara and H. Toki, Nucl. Phys. \textbf{A579}, 557 (1994).

\bibitem{IUFSU} F. J. Fattoyev, C. J. Horowitz, J. Piekarewicz, and G. Shen,
Phys. Rev. C \textbf{82}, 055803 (2010).

\bibitem{Shen02} H. Shen, Phys. Rev. C \textbf{65}, 035802 (2002).

\bibitem{Shen11} H. Shen, H. Toki, K. Oyamatsu, K. Sumiyoshi,
Astrophys. J. Suppl. \textbf{197}, 20 (2011).

\bibitem{Horo01} C. J. Horowitz and J. Piekarewicz,
Phys. Rev. Lett. \textbf{86}, 5647 (2001).

\bibitem{Horo03} J. Carriere, C. J. Horowitz, and J. Piekarewicz,
Astrophys. J. \textbf{593}, 463 (2003).

\bibitem{Prov13} C. Provid\^{e}ncia and A. Rabhi, Phys. Rev. C \textbf{87}, 055801 (2013).

\bibitem{Maru05} T. Maruyama, T. Tatsumi, D. N. Voskresensky, T. Tanigawa, and S. Chiba,
Rhys. Rev. C \textbf{72}, 015802 (2005).

\bibitem{SSA10} S. S. Avancini, C. C. Barros Jr., D. P. Menezes, and C. Provid\^{e}ncia,
Phys. Rev. C \textbf{82}, 025808 (2010).

\bibitem{SSA09} S. S. Avancini, L. Brito, J. R. Marinelli, D. P. Menezes, M. M. W. deMoraes,
C. Provid\^{e}ncia, and A. M. Santos, Phys. Rev. C \textbf{79}, 035804 (2009).

\bibitem{Cent98} M. Centelles, M. Del Estal, and X. Vi\~{n}as, Nucl. Phys. \textbf{A635}, 193 (1998).

\bibitem{Douc00} F. Douchin, P. Haensel, and J. Meyer, Nucl. Phys. \textbf{A665}, 419 (2000).

\bibitem{Demo10} P. B. Demorest, T. Pennucci, S. M. Ransom, M. S. E. Roberts, and J. W. T. Hessels,
Nature (London) \textbf{467}, 1081 (2010).

\bibitem{Duco10} C. Ducoin, J. Margueron, and C. Provid\^{e}ncia,
Europhys. Lett. \textbf{91}, 32001 (2010).

\end{thebibliography}
\end{document}